\documentclass[sigconf]{acmart}

\AtBeginDocument{%
  \providecommand\BibTeX{{%
    \normalfont B\kern-0.5em{\scshape i\kern-0.25em b}\kern-0.8em\TeX}}}

\copyrightyear{2019} 
\acmYear{2019} 
\setcopyright{rightsretained} 
\acmConference[ICPP 2019]{48th International Conference on Parallel Processing}{August 5--8, 2019}{Kyoto, Japan}
\acmBooktitle{48th International Conference on Parallel Processing (ICPP 2019), August 5--8, 2019, Kyoto, Japan}\acmDOI{10.1145/3337821.3337839}
\acmISBN{978-1-4503-6295-5/19/08}

\usepackage[utf8]{inputenc} 
\usepackage[T1]{fontenc}    
\usepackage{hyperref}       
\usepackage{url}            
\usepackage{booktabs}       
\usepackage{spreadtab}      
\usepackage{amsfonts}       
\usepackage{nicefrac}       
\usepackage{microtype}      
\usepackage{color,soul}
\usepackage{subcaption}
\usepackage{graphicx}
\usepackage{comment}
\DeclareMathAlphabet{\mathcal}{OMS}{cmsy}{m}{n}
\SetMathAlphabet{\mathcal}{bold}{OMS}{cmsy}{b}{n}
\usepackage{multirow}

\begin{document}


%
\title{A Unified Optimization Approach for CNN Model Inference on Integrated GPUs}

%
\author{Leyuan Wang}
\email{wangleyu@amazon.com}
\affiliation{%
  \institution{Amazon Web Services}
  \city{East Palo Alto}
  \state{CA}
  \country{USA} 
}

\author{Zhi Chen}
\email{chzhi@amazon.com}
\affiliation{%
  \institution{Amazon Web Services}
  \city{East Palo Alto}
  \state{CA}
  \country{USA} 
}

\author{Yizhi Liu}
\email{yizhiliu@amazon.com}
\affiliation{%
  \institution{Amazon Web Services}
  \city{East Palo Alto}
  \state{CA}
  \country{USA} 
}

\author{Yao Wang}
\email{wayao@amazon.com}
\affiliation{%
  \institution{Amazon Web Services}
  \city{East Palo Alto}
  \state{CA}
  \country{USA} 
}

\author{Lianmin Zheng}
\email{lianminzheng@gmail.com}
\affiliation{%
  \institution{Shanghai Jiaotong University}
  \city{Shanghai}
  \country{China} 
}

\author{Mu Li}
\email{mli@amazon.com}
\affiliation{%
  \institution{Amazon Web Services}
  \city{East Palo Alto}
  \state{CA}
  \country{USA} 
}

\author{Yida Wang}
\email{wangyida@amazon.com}
\affiliation{%
  \institution{Amazon Web Services}
  \city{East Palo Alto}
  \state{CA}
  \country{USA} 
}

%
\renewcommand{\shortauthors}{Wang, et al.}

%
\begin{abstract}
Modern deep learning applications urge to push the model inference taking place at the edge devices for multiple reasons such as achieving shorter latency, relieving the burden of the network connecting to the cloud, and protecting user privacy. The Convolutional Neural Network (\emph{CNN}) is one of the most widely used model family in the applications. Given the high computational complexity of the CNN models, it is favorable to execute them on the integrated GPUs at the edge devices, which are ubiquitous and have more power and better energy efficiency than the accompanying CPUs. However, programming on integrated GPUs efficiently is challenging due to the variety of their architectures and programming interfaces. This paper proposes an end-to-end solution to execute CNN model inference on the integrated GPUs at the edge, which uses a unified IR to represent and optimize vision-specific operators on integrated GPUs from multiple vendors, as well as leverages machine learning-based scheduling search schemes to optimize computationally-intensive operators like convolution. Our solution even provides a fallback mechanism for operators not suitable or convenient to run on GPUs. The evaluation results suggest that compared to state-of-the-art solutions backed up by the vendor-provided high-performance libraries on Intel Graphics, ARM Mali GPU, and Nvidia integrated Maxwell GPU, our solution achieves similar, or even better (up to 1.62$\times$), performance on a number of popular image classification and object detection models. In addition, our solution has a wider model coverage and is more flexible to embrace new models. Our solution has been adopted in production services in AWS and is open-sourced.
\end{abstract}

%
%

%

%

\maketitle
\section{Introduction}
\label{sec:intro}
The recent advance of deep learning enables a number of sophisticated applications taking place at the edge, making the prevailing edge devices, such as camera, speaker, television, and mobile phone, around us smart. These applications, ranging from computer vision related tasks, such as image classification, object detection, and segmentation, to speech recognition and voice detection, typically leverage pre-trained deep learning models to do inference on the input data. It is more and more critical to execute the model inference directly at the edge devices for shorter latency, less burden of the network bandwidth, and better privacy protection to the users.

A typical edge device equips with an SoC (System on a Chip), which integrates various compute units such as CPU, GPU, and optionally DSP (Digital System Processor) and NPU (Neural Processing Unit). The integrated GPUs, although normally much less powerful than server side discrete GPUs, are able to deliver higher FLOPs than the accompanying CPUs. However, in practice, the majority of model inference at the edge is executed on CPUs~\cite{wu2019hpca} because of easier programmability and more flexible portability across different SoCs.

It is actually less favorable to execute deep learning model inferences on CPUs at the edge due to two reasons. First, CPU is normally less powerful than the integrated GPU located in the same SoC. For example, in the three platforms that we used for experiments in Section~\ref{sec:eval}, the theoretical peak FLOPs of GPUs are 5.16$\times$, 6.77$\times$, and 2.48$\times$ greater than the accompanying CPUs, on AWS DeepLens (Intel HD 505), Acer aiSage (ARM Mali T-860), and Nvidia Jetson Nano (Nvidia GPU with Maxell architecture), respectively. In addition, CPUs may suffer from power throttling when overheated, leading to dramatically reduced performance. Second, the execution time on CPUs is less stable compared to GPUs. Besides the possible power throttling, the operating system of the device normally has multiple processes (e.g. daemons) periodically running on CPUs which inevitably causes CPU resource contention and consequently high variance of the model inference duration~\cite{kayiran2014micro}.

In practice, integrated GPUs observe limited usage in deep learning model inference due to the lack of a generic solution. The integrated GPUs in different SoCs from different chip vendors vary vastly. For example, there is a wide disparity between Intel Graphics and ARM Mali GPUs in terms of architecture. Intel Graphics has a \emph{subgroup} concept when organizing the threads. The running threads within the same \emph{subgroup} share the same register file. Thus, sharing data within a \emph{subgroup} could boost the performance on Intel Graphics significantly. However, \emph{subgroup} is not present in ARM Mali GPUs, hence the optimization customized for Intel Graphics is not applicable to devices equipped with ARM Mali GPUs. In addition, different GPUs may require different software drivers and programming languages, such as CUDA on Nvidia products and OpenCL on Intel and ARM products. It is difficult for developers to transfer the optimization solutions from one hardware platform to another, let alone guaranteeing the reasonable performance.

In order to provide GPU support, practitioners usually leverage the chip vendor provided high-performance libraries, e.g. Intel clDNN~\cite{cldnn}, ARM Compute Library (\emph{ACL})~\cite{acl}, and Nvidia cuDNN~\cite{chetlur2014cudnn}, along with some deep learning frameworks or vendor-provided inference-only pipelines (Intel OpenVINO~\cite{openvino} and e.g. Nvidia TensorRT~\cite{tensorrt}), to build up their applications. While it is likely to deliver reasonable performance, this solution is known to be inflexible and error-prone, plus requiring a large amount of tedious engineering efforts. Furthermore, these libraries focus heavily on the computationally-intensive operators like convolution, while leaving some vision-specific operators to run on CPU (e.g. Non Maximum Suppression (\emph{NMS})) or on GPU with suboptimal performance (e.g. \emph{ROIAlign}). The vision-specific operators normally require much less computation than convolution, but they involve non-straightforward control logic which is not in the GPU-favored computation style. In addition, the input length of these operators could be notably greater than the number of processors of the GPU, hence synchronizations is needed. Meanwhile, binding the applications to the third-party libraries greatly delays software development cycles. For example, developers have to wait for the vendor libraries to support the new features for deploying the new models into production when new deep learning models are invented and new operators are introduced. As we will show in the evaluation, many start-of-the-art CNN models lack off-the-shelf optimized implementation from vendor libraries.

To surmount these limitations, this paper proposes a unified end-to-end stack to deploy and optimize CNN models for efficient inference on mainstream integrated GPUs from Intel, ARM and Nvidia. Note that our proposed solution is not limited to edge GPUs but applies to cloud GPUs as well, which is beyond the scope of this paper. We choose CNN models as they are heavily used in image and video processing tasks which represent the primary use cases at edge devices. Experimental results show that our solution runs efficiently on the aforementioned integrated GPUs, achieving similar, or even better (up to 1.62$\times$) performance compared to the vendor-provided solutions on five state-of-the-art image classification and object detection model families. In addition, our solution has wider model coverage. Our work is already deployed in Amazon SageMaker Neo Service~\footnote{\url{https://aws.amazon.com/sagemaker/neo/}}, enabling model developers to optimize for inference at the edge. Using this service, a number of application developers have deployed CNN models optimized for inference in production on several types of edge devices. The code of this project is also open-sourced as part of the TVM stack~\footnote{\url{https://github.com/dmlc/tvm/}}.

To summarize, our paper presents a solid engineering work with the following contributions:
\begin{enumerate}
\item An open-source end-to-end system that runs a wide range of CNN models for image classification and object detection on integrated GPUs;
\item A thorough investigation on optimizing vision-specific operators as well as the computationally-intensive tensor operators on integrated GPUs;
\item A systematical empirical evaluation of a number of CNN models on multiple integrated GPUs. 
\end{enumerate}

The rest of the paper is organized as following. Section~\ref{sec:background} goes over the background of our work, including the architecture and optimization schemes of the mainstream integrated GPUs from Intel, ARM and Nvidia, and the typical CNN model inference workloads on these GPUs and the existing unified optimization approach for these workloads. We propose the methodology of our end-to-end software stack to run CNN model inference efficiently on integrated GPUs in Section~\ref{sec:opt}. Section~\ref{sec:eval} summarizes the evaluation results of our solution on three different platforms featured with Intel, ARM and Nvidia integrated GPUs. Section~\ref{sec:related} lists the related works, followed by Section~\ref{sec:concl} that concludes the paper.
\section{Background}
\label{sec:background}
\subsection{Integrated GPUs}
\label{sec:background:gpu}
This paper talks about optimizing CNN model inference on integrated GPUs at edge devices. The mainstream integrated GPUs are produced by Intel, ARM and Nvidia. These GPUs are on-die connecting to other agents within the same SoC like CPU cores via a ring interconnect and they share the main memory with CPU cores. The integrated GPUs maintain several levels of hierarchical caches to reduce data latency, typically including register files, L1 and L2 caches. An efficient computation pattern will mostly use the data stored in the register files and hide the latency to retrieve data from farther memories.

The integrated GPUs process the computation using their compute units, which are called \emph{execution units (EUs)} in Intel graphics, \emph{shader cores (SCs)} in ARM Mali GPUs, and \emph{stream multiprocessors (SMs)} in Nvidia GPUs. Each compute unit coordinates a certain number of hardware threads and each thread owns a certain amount of register files. According to the generation and the level of the integrated GPU, the number of compute units varies. Most of the modern integrated GPUs support SIMD instructions in the algorithmic ISAs. An efficient computation pattern should keep all the available threads of all compute units busy for most of the time and utilize the SIMD instructions whenever is possible.

Integrated GPUs target two similar programming models for general-purpose programmability,
OpenCL~\cite{Howes:2014:TOS} (managed by Khronos) and CUDA~\cite{NVIDIA:2014:CUDA} (developed by Nvidia).
Integrated GPUs on Intel and ARM devices normally support OpenCL as the driver and utilize its APIs to program, which is the most general-purpose program to run on these GPUs. On the other hand, Nvidia products run CUDA as its proprietary driver which outperforms OpenCL~\cite{karimi2010performance}. Despite significant difference in details, OpenCL and CUDA actually share similar abstractions.

The modern integrated GPU is a massively parallel processor that supports hundreds of hardware scheduled threads running simultaneously. These threads are organized into blocks (OpenCL: workgroups) and the hardware schedules blocks of threads onto hardware cores (CUDA: streaming multiprocessors, OpenCL: compute units). Nvidia GPUs have on the order of 16 cores, each of which contains 32-wide SIMD processors (CUDA: CUDA cores, OpenCL: SIMD units) that run 32 threads in lockstep. GPUs also feature a memory hierarchy of per-thread registers, per-block shared memory (per-work-group local memory), and off-chip global DRAM accessible to all threads. CUDA programs (``kernels'') specify the number of blocks and threads per block under a SIMT (single-instruction, multiple-thread) programming model. GPU implementations typically launch a number of kernels during execution. A kernel is essentially a function, which can be instantiated into many instances to deal with different data specified by block indices. They achieve parallelization by running many kernel instances simultaneously, each of which is called a \emph{work item}. In integrated GPUs, a work item corresponds to a SIMD entry, processed by a CUDA core (CUDA term) or a virtual thread (OpenCL term in Intel Graphics). Therefore, a warp (CUDA term) or hardware thread (OpenCL term in Intel Graphics) processes multiple work items at the same time, inherently implementing the SIMD vectorization. This programming model of CUDA and OpenCL is suitable for the compute pattern of neural networks. Efficient GPU programs should 1) have enough work per kernel to keep all hardware cores busy (load balancing); 2) strive to reduce thread divergence (when neighboring threads branch in different directions); 3) aim to access memory in large contiguous chunks to maximize achieved memory bandwidth (coalescing); and 4) minimize communication between CPU and GPU. Designing an implementation that achieves all of these goals is a significant challenge. 

In order to utilize integrated GPUs well, the chip vendors normally ship products with well-optimized high-performance libraries, e.g. clDNN for Intel graphics, ACL for ARM Mali GPUs, and cuDNN for Nvidia GPUs. These libraries take advantage of properties of CUDA/OpenCL described above to achieve good performance on GPUs. In addition, Intel extends the OpenCL driver to support some special features of their hardware platforms. For instance, Intel-extended OpenCL organizes the work items of the same hardware thread as a \emph{subgroup}, which share the same register files of the hardware threads. These high-performance libraries target mostly on optimizing the computationally-intensive tensor operators like convolution. 

In order to efficiently execute the entire model inference workloads, Nvidia and Intel wrap their high-performance libraries as TensorRT and OpenVINO, respectively. However, neither TensorRT nor OpenVINO is open-sourced, making it inflexible for developers customize them for new or slightly changed models. Regarding ARM Mali GPUs, we are not aware of an ARM-provided model inference solution. Instead, practitioners use deep learning frameworks which incorporate ACL to seek for good performance, which normally is not reliable and lacks coverage.

\subsection{CNN Model Inference at the Edge}
\label{sec:background:cnn}
We are observing more and more applications at edge devices including image and video processing tasks, such as image classification, object detection, and segmentation. As we discussed in Section~\ref{sec:intro}, these tasks are done by CNN model inference, which is preferably executed on the integrated GPUs at the edge devices.

However, it is challenging to efficiently fulfill this task due to two reasons. First, CNN models consist of a large number of computationally-intensive convolution operations. Fully Optimizing them for integrated GPUs is non-trivial. On one hand, the architecture of the integrated GPUs varies vastly between vendors; that is, the optimized solution on one GPU may not be applicable to others at all. On the other hand, convolutions with different data input shapes may require different optimization schemes. There is no panacea to all possible convolution workloads. Therefore, the optimization on convolutions should be conducted case by case.

Second, there are some vision-specific operators in object detection and segmentation models, such as \emph{NMS} and \emph{ROIAlign}. These operators typically do not require intensive computation but require control-flow logic that GPUs are not naturally suitable to handle. For example, SSD yields a large number of predictions to achieve more coverage of location, scale, and aspect ratios. The \emph{NMS} operator then scans these predictions to remove duplicates that point to the same object and finally sorts the updated predictions by confidence scores. For this type of operators, normally there is no high-performance implementation, or even no implementation on the integrated GPUs (such as Intel Graphics). It severely blocks the deployment of the corresponding models for efficient execution.

\subsection{Existing unified optimization solution}
\label{sec:background:tvm}
Our solution is built on top of the open-source deep learning compiler stack TVM~\cite{chen2018tvm}. It was designed for compiling and optimizing deep learning model inference across multiple hardware platforms including CPUs, GPUs, and specialized accelerators. Specifically, inherited from Halide~\cite{ragan2013halide}, TVM utilizes a unified IR to lower the optimization schemes of different integrated GPUs to CUDA or OpenCL for code generation on those devices. However, a number of commonly-used CNN models, e.g. SSD~\cite{liu2016ssd} and Yolo~\cite{redmon2018yolov3}, are not fully supported due to the missing of some vision-specific operators. Even for those models whose operators are covered by TVM, the end-to-end performance of CNN model inference on the mainstream integrated GPUs is generally not appealing because the control logic-involved operators are not carefully tuned for GPUs, and the scheduling schemes for computationally-intensive tensor operators are not thoroughly explored.

Our solution extends the original TVM in multiple ways. For the vision-specific operators that require non-straightforward control flow logic, we implemented them via the unified IR of TVM in an optimized fashion, which thoroughly utilizes the available computing resources of the GPU and is applicable to integrated GPUs provided by different vendors. For the computationally-intensive operators like convolution, we resort to the machine learning-based approach~\cite{Chen:2018:LOT, liu2018optimizing} to automatically search good optimization schemes for different convolution workloads on different GPUs. In addition, in order to facilitate operators to fall back to CPUs for easier execution, we enabled the heterogeneous execution to the original TVM stack.
\section{Methodology}
\label{sec:opt}
\begin{figure}[t]
	\centering
  \includegraphics[width=1.0\columnwidth]{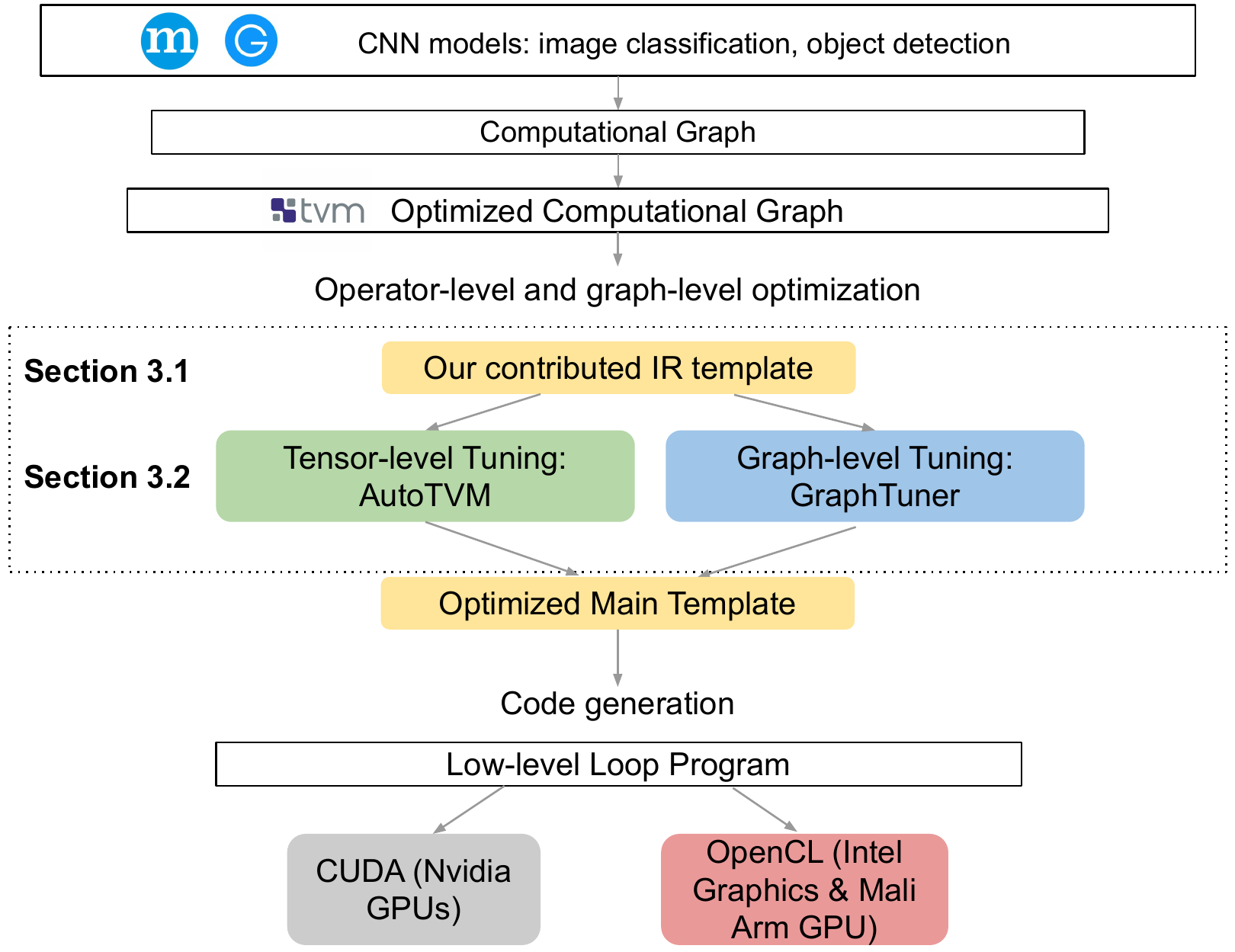}
	\caption{Overview of our working pipeline. Note that the universal GPU IR generated by our solution works for both CUDA and OpenCL.}
	\label{fig:overview}
\end{figure}

This section focuses on the methodology that we used to obtain the high performance and wide model coverage using a unified optimization approach. Our optimization consists of two parts. First, we will discuss how to optimize the vision-specific operators on integrated GPUs, which is non-trivial but largely ignored in the existing work. Section~\ref{sec:opt:cv} covers the GPU optimization and Section~\ref{sec:opt:fallback} provides an alternative way to fall back the operators that are not GPU-friendly to CPUs if needed. Second, we present the insights of attaining performance gains on computationally-intensive tensor operators like convolution. Section~\ref{sec:opt:opencl} first goes over the optimization strategies of convolution on Intel Graphics, which is not well-studied in the literature and then discusses two types of machine learning-based techniques to explore the optimization space for achieving better performance. As described in Section~\ref{sec:background:tvm}, our solution is based on TVM~\cite{Chen:2018:LOT} with a number of improvements. An overview of the working pipeline is shown in Figure~\ref{fig:overview}.

\subsection{Vision-specific Operators}
\subsubsection{Optimization for Integrated GPUs}
\label{sec:opt:cv}
Object detection models are popularly used in the applications at the edge. These models extend from CNN-based image classification models by adding a number of post-classification, vision-specific operators. These operators are normally used to propose regions of interest and sort them accordingly. Vision-specific operators do not require much computation but are demanding to optimize on integrated GPUs for performance concern. This is because these operators usually involve non-straightforward control logic which requires the threads within a compute unit of a GPU to diverge if not handled carefully. As a result, it is practically difficult to run object detection models, such as SSD and Yolo, entirely on integrated GPUs. Current solutions can at most run in sub-optimal performance. This subsection describes how we optimize the vision-specific operators that essentially cause the lack of support of CNN models on integrated GPUs.

\paragraph{Sorting}
\label{sort}
\emph{Sorting}, essentially, \emph{argsort}, is a common operator in CNN models, such as SSD~\cite{liu2016ssd}. Basically, \emph{argsort} assigns the sorted index number to every item in the list~\footnote{\url{https://docs.scipy.org/doc/numpy/reference/generated/numpy.argsort.html}}. There are fast and efficient \emph{argsort} implementations for discrete GPUs in CUDA~\cite{CUB:2016, MGPU:2016}. However, the OpenCL counterpart for integrated GPUs of Intel and ARM is not yet available. Achieving high performance sorting on GPUs is notoriously hard as it requires conditional branches that do not fit the parallel nature of GPUs. For example, the \emph{NMS} operators frequently used in SSD models contain sorting operations to sort the small data blocks where each of them may vary in the input size. In other words, for each dimension of the input data, different numbers of elements need to be sorted. This process could cause branch divergence if not implemented properly. GPUs are not designed to run efficiently on small imbalanced problems that are substantially harmful to the performance.

\begin{figure}[t]
	\centering
  \includegraphics[width=0.808\columnwidth]{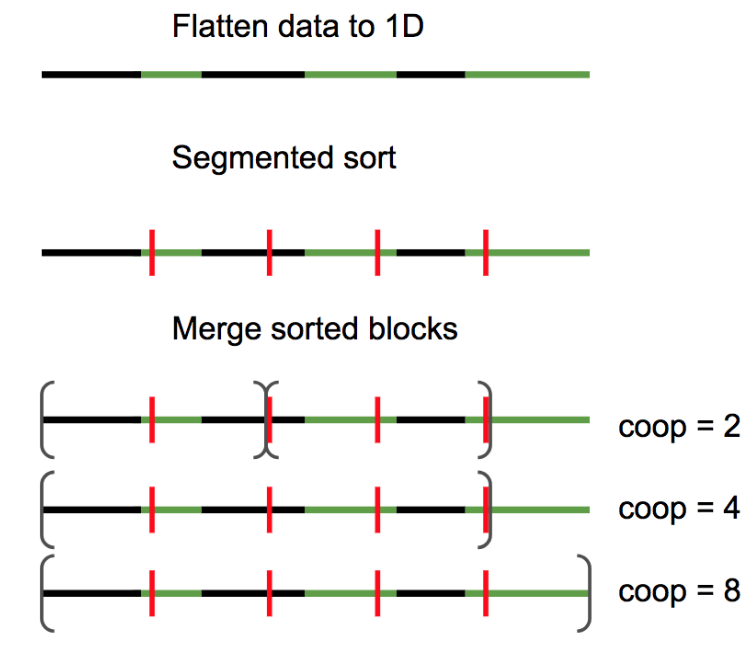}
	\caption{Segmented sort pipeline. }
	\label{fig:argsort}
\vspace{-0.2in}
\end{figure}

To tackle this issue, we proposed to flatten the input array so that the fine-grained sorting problem at each dimension becomes coarse-grained. This, however, brings us two challenges as well, load balancing and programming simplicity. We used segmented sorting to optimize the \emph{argsort} for CNN models as illustrated in Figure~\ref{fig:argsort}. Note that although our implementation uses similar ideas as in previous segmented sort work~\cite{Hou:2017:FSS} and merge sort work ~\cite{Wang:2016:FPS}, it is not limited to Nvidia GPUs but uses the unified program to run efficiently on different integrated GPUs with different architectures.

The idea is as follows. First, the data is flattened into an array with the starting index of each segment stored. Second, the flattened array is divided into equal length, instead of sorting each individual variant length segment. Third, block sorting is accomplished on each segment to maintain their ordering. Finally, we perform a series of merge operations. Each iteration doubles the cooperative block size until all elements are sorted. For example, in Figure~\ref{fig:argsort}, black and green lines represent different segments. Length of the lines represents the size of a segment. Red vertical lines represent active interface. We chop the flattened array into equal-size blocks. For ease of illustration, we assume that there are five thread cores in total. Each thread operates on one block. So after doing local sorting inside each block, we do merge sort. In coop 2, 2 threads work cooperatively to merge two blocks. In coop 4, four threads work together to merge four blocks. And in coop 8, all five blocks are merged by all thread cores working cooperatively. and  This algorithm is efficient because only the segments that span the active interface between two input lists are modified.
\paragraph{Prefix Sum}
\label{scan}
\emph{Prefix sum (Scan)} is a trivial sequential algorithm on the CPU. But GPU is not optimized for doing sequential computations especially when work items have data dependence. Many algorithms in the literature~\cite{Harris:2007:PPS, Owens:2008:GC, Sengupta:2007:SPF} have optimized scan on Nvidia GPUs (CUDA specific) but none of them targeted all integrated GPUs. Based on the classic parallel scan by Hillis and Steele~\cite{Hillis:1986:DPA}, we proposed an efficient prefix sum for both CUDA- and OpenCL-oriented integrated GPUs.

\begin{figure}[t]
	\centering
  \includegraphics[width=\columnwidth]{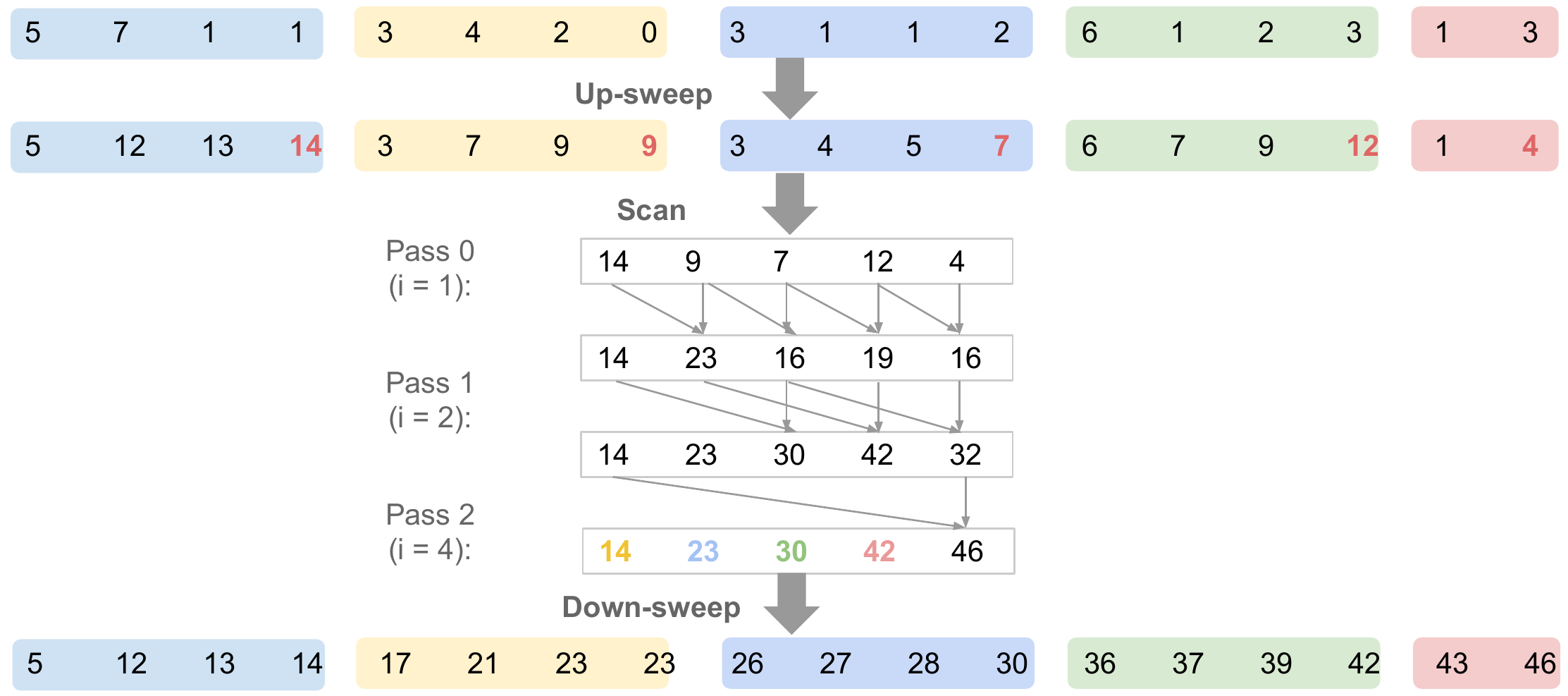}
	\caption{Prefix sum (Scan) pipeline example. Suppose we have 5 parallel processors.}
	\label{fig:scan}
\end{figure}

Our solution was designed in a three-stage manner: \textit{up-sweep}, \textit{scan}, and \textit{down-sweep}. Given $n$ input elements, $\log{}n$ passes are needed to complete the scan in the cooperative scan of the Hillis and Steele's method. In pass $d$, the element $i - 2^d$ is added to the element $i$. Though the algorithm has $\mathcal{O}(n\log n)$ operations compared to $\mathcal{O}(n)$ for the CPU sequential algorithm, it reduces the latency from $\mathcal{O}(n)$ to $\mathcal{O}(\log n)$. However, the number of inputs is normally much larger than the number of cores in a device. Therefore, simply applying the previously mentioned method is inefficient as it may trigger global synchronization for each pass. We leverage register blocking to avoid the need of global synchronization and reduce data movement. Register blocking technique assigns multiple elements to one processor which are processed sequentially. The reduction results from all cores are then processed with a parallel scan using Hillis and Steele's method without global synchronization. Fig~\ref{fig:scan} illustrates an example of our solution. Suppose there are 5 parallel processors and 18 elements sitting in an array, we assign 4 elements to each processor except the last one. For \textit{up-sweep} step, scan computation is done sequentially inside each processor (color block) and parallel across all processors. After getting segmented reduction results for each processor (numbers in red bold), we do parallel scan across all processors. For \textit{down-sweep} step, the scanned results are added back to the corresponding processors in parallel (color of number matching color of the processor block).
\paragraph{Other Vision-specific Operators}
\label{cv_op}
There are some other vision-specific operators, such as $box\_nms$, $ROIAlign$, $MultiboxDetection$, known to be tedious to implement in integrated GPUs in an efficient way. Our solution leverages TVM IR to produce fairly fast GPU code with less demanding engineering efforts. For example, our approach only requires around 100 lines of TVM IR code (vs 325 lines of CUDA code in the original implementation) to generate efficient code for both CUDA and OpenCL supported platforms.

It turned out that the implementation of all vision-specific operators for targeted models on integrated GPUs is still very time consuming though we have managed to do so. Actually, the common practice of handling not-GPU-friendly operators in SoC is to fall back these operators to CPUs. The next section discusses how We enabled this functionality.

\subsubsection{Fallback}
\label{sec:opt:fallback}
While integrated GPUs have shown their effectiveness in producing competitive performance with reasonable power budget, there are some circumstances that we may not be able to fully execute the whole model on integrated GPUs. First, GPUs generally offer less flexibility and programmability than their accompanying CPUs. Hence, some vision-specific operators, e.g. sorting, that require intensive control logic are intrinsically more difficult to implement on GPUs than on CPUs. As discussed in Section~\ref{sec:opt:cv}, one may spend more engineering efforts to write the GPU versions of these operators for each GPU vendor and tune performance for them. A more effective alternative is to fall back these operators to CPUs for much simpler and more unified implementation, but leave the other computationally-intensive parts running on GPUs. It is a reasonable design due to the following three reasons. 

First, this kind of operators typically do not require much computation and hence no massive parallelism to exploit. Second, these operators are usually found in the pre/post-processing section of a neural network which is not the performance-critical part, so executing them on CPUs does not bring much performance penalty. Third, in practice, the amount of data transmission required to fall back is not large due to the above reasons and there is not much back-and-forth data movement/copy across GPU and CPU devices. In addition, the wide use of shared memory has facilitated the data transferring between GPU and CPU on the same SoC. Recent Intel Graphics processor even features dedicated shared local memory to enable the sharing of programmer managed data among kernel instances within an OpenCL work group.

It may seem that a sophisticated algorithm is required to intelligently place the right operator on the right device. However, it turned out that a simple heuristic is sufficient for CNN models as we attempt to schedule as many operators on the integrated GPUs, only leaving a few GPU unfriendly ones to the CPU. This heuristic is implemented using a standard graph traversal technique in a two-pass manner with a list of know operators that are performant on GPUs. In the first pass, we tag the device property of a graph node as GPU once it is matched in the list. Otherwise, its device property is labeled as CPU. On the completion of the first pass, the second pass is carried out to insert a data copy operator between any two directly connected nodes assigned to different devices. Our empirical results verified that the fallback approach only causes negligible performance degradation. For example, running SSD model inference (backed with ResNet) of one sample entirely on the integrated GPU of AWS DeepLens took 1010.23 ms, while falling back the \emph{nms} operators which involves sorting to CPU took 1015.14 ms, leading to an overhead less than 0.5\%. It is worth noting that in the experiment we ran the entire model inference on the integrated GPUs. However, our fallback mechanism enables the possible early adoption of new models with new operators.

\subsection{Computationally-intensive Operators}
\label{sec:opt:opencl}
\subsubsection{Optimization Consideration on Intel Graphics}
While integrated GPUs from ARM and Nvidia have been gathered a number of research interests recently~\cite{marti2019hpca, Grasso2014ipdps, chetlur2014cudnn}, it is challenging to extend the successful optimization tricks to Intel Graphics. To the best of our knowledge, there is virtually no published work that provides detailed insights regarding the methodology of performance optimization on Intel Graphics for various CNN models. As the main memory and last-level cache of Intel Graphics are treated as the global memory and register files and L1 to L3 caches are local, an intuitive solution to boost performance is to store as much as possible data in the local memory. However, that this scheme usually does not lead to the expected results because the closest memory, namely, the \emph{general-purpose register files (GRFs)}, is playing a much more critical role than others. If they are not properly used, even if data are kept in the local caches, the performance would not be appealing. The GRFs could be effectively employed by an execution unit to perform SIMD computations. More importantly, the SIMD floating point unit of an execution unit is able to combine multiple 128-bit registers together to form much wider registers so that up to eight 32-bit floating point values can be computed in parallel. Allocating the work to the Intel Graphics in the unit of OpenCL kernels is a plausible means to maximize the utilization of the SIMD units that leverage the register files efficiently.

As discussed in Section~\ref{sec:background:gpu}, for Intel Graphics, a hardware thread consists of multiple virtual threads building up a \emph{subgroup} that shares this hardware thread's 4KB GRFs. This extension is designed to allow work items in a subgroup to share data without the use of local memory and work group barriers, and to utilize specialized hardware to load and store blocks of data. In the Intel OpenCL extension, there are primitives, such as $intel\_subgroup block\_read$, $intel\_subgroup\_block\_write$ and $intel\_subgroup\_shuffle$, that enable the data reading/writing in the GRFs and broadcasting across the work items of a hardware thread. Multiple subgroups are composed as a group, in which the subgroups are processed simultaneously using multiple hardware threads. An OpenCL program typically contains multiple groups. OpenCL organizes the groups and virtual threads in up to three dimensions. The virtual threads belonging to the same Intel-defined subgroup should be in the same dimension within a group.

By taking into account the aforementioned characteristics of OpenCL on Intel Graphics, we proposed heuristics to optimize the deep learning operators on these GPUs.

\subsubsection{Convolution Optimization}
\label{sec:opt:opencl:ops}
For CNN models, \emph{Conv2D} is the most time-consuming operator that we need to focus on. As an overview of the pipeline, we wrote an optimized schedule template (Section~\ref{sec:opt:opencl:ops}), and then used AutoTVM~\cite{Chen:2018:LOT} as well as graph tuner~\cite{liu2018optimizing} to search the best schedules for different workloads (section~\ref{sec:opt:opencl:tuning}). We adaptively adjust the main template and revamp the search whenever there is a headroom for performance improvement.

As in~\cite{liu2018optimizing}, we used spatial packing, namely loop tiling, to maximally reuse memory during the convolution computation. Loop tiling splits a loop's iteration space into smaller blocks to make sure data used in a loop remains in the cache when it is reused. A loop nest might be partitioned into a number of smaller blocks with a reasonable tiling size. Each of these partitioned blocks is able to perfectly fit into the cache so as to fully take advantage of spatial locality. For Intel Graphics, we also fully utilize the advantage of \textit{subgroup} by arranging the convolution kernels to fit into the GRFs of a subgroup.

In sum, we propose the following heuristics for convolution operators:
\begin{itemize}
    \item The output channels are divided so that they can be parallelized in multiple groups;
    \item Splitting the feature map along the height dimension so that multiple groups can be executed in parallel;
    \item Unrolling the nested loops of a convolution kernel. Loop unrolling normally attains benefits including reduced control overhead (by removing testing exit conditions), increased instruction level parallelism, better opportunity of scalar optimizations such as constant folding, and/or more friendly code for (partial) vectorization~\cite{Gong2018oospla}.
\end{itemize}

From the above heuristics one can find that there are many parameters whose values depend on different workloads. The shape of the work groups significantly matters since it controls the memory reuse among threads. The tile size is also performance sensitive as it affects the reuse of data at a specific level of the memory hierarchy. It is difficult to determine ``the best'' combination of parameters for a specific workload on a specific platform as the number of workloads and platforms are both large. In the next section, we will introduce machine learning-based heuristics to search for the most efficient combination of these parameters in an automatic way to enhance performance for convolution kernels. On top of it, we will also bring in more optimization approaches in the computational graph level.

\subsubsection{Performance Tuning}
\label{sec:opt:opencl:tuning}
Based on the optimization guidance provided in the last section, one could build up a large optimization search space using different parameters. These parameters could be from either graph- or tensor-level. Therefore, we are able to tune the performance at different levels of the representations of a given CNN model.

\paragraph{Graph-level tuning: Graph Tuner}
Graph-level optimization has great impact on CNN model performance. Besides general graph-level optimizations, including operator fusion, pre-computing, simplifying inference for batch-norm and dropout~\cite{chen2018tvm}, we also applied the graph tuner technique proposed in~\cite{liu2018optimizing} to fine-tune the data layout to achieve better end-to-end performance. Since optimizing convolution kernels requires transforming input and output to different data layouts which might bring extra overhead, the graph tuner uses dynamic programming to examine the trade-off between optimized kernels and data layout transformation overheads. By applying all these graph-level optimization strategies, we are able to get the graph-level optimal schedule for each convolution kernel. To the best our knowledge, this is the first work to introduce the thorough graph-level tuning for CNN model inference optimization on integrated GPUs.

\paragraph{Tensor-level tuning: AutoTVM}
CNN operators normally act on top of tensors (e.g. n-dimensional arrays). Hand written schedule/optimization libraries, e.g. cuDNN, MKL-DNN, etc, from hardware vendors have been widely used to deliver compelling performance. However, each vendor may provide their own proprietary library with optimized performance on a certain number of kernels. That is, programmers may have to sort out the best combination of various kernel implementations. It requires significant engineering efforts and domain knowledge of each kernel from each vendor. AutoTVM~\cite{Chen:2018:LOT} provides a venue that automatically optimizes the common tensor operators for given hardware and builds up an optimization space composing possible transformed versions of tensors. Both hardware and software related factors are considered when constructing a low-level transformed program. For instance, the typical ones are unrolling factor, thread binding, and vectorization capabilities, etc. Users then only have to provide limited parameters (a.k.a templates) to explore the search space based on statistical cost models for predicting achievable performance results. In our solution, the AutoTVM tool is utilized to search for scheduling schemes that lead to decent performance of convolutions based on the templates we constructed in Section~\ref{sec:opt:opencl:ops}. Although AutoTVM has been used on ARM Mali GPUs~\cite{Chen:2018:LOT}, we conduct a more comprehensive study on applying it to a number of mainstream integrated GPUs for a wide range of CNN models. It is worth noting that doing tensor-level search is costly particularly at the edge devices due to their limited compute capability. For instance, it took up to tens of hours to search all convolution workloads in one model for one device. In order to prevent replicated searching in the future, we maintain a database to store the results for every convolution workload on each hardware platform.

\section{Evaluation}
\label{sec:eval}
This section seeks to answer the following questions.
\begin{enumerate}
\item What is the coverage and overall performance of our end-to-end solution compared with the state-of-the-art alternatives for multiple commonly used computer vision models on popular edge devices?
\item What is the performance boost we achieved via optimizing the vision-specific operators for integrated GPUs?
\item What is the performance boost we achieved by applying the machine learning-based scheduling search?
\end{enumerate}

\subsection{Experiment Setup}
\label{sec:eval:setup}
We evaluated our approach on three representative edge devices, AWS DeepLens, Acer aiSage, and Nvidia Jetson Nano. AWS DeepLens is a smart video camera and development platform equipped with an Intel Atom processor x5-E3930 SoC whose integrated GPU is Intel Graphics HD 505 (HD series in Gen 9th). Acer aiSage is an edge computing device that combines AI and computer vision to process image/video data from digital TVs, which is powered by an RK3399 SoC including ARM Mali GPU T-860 (Midgard 4th Gen). Nvidia Jetson Nano, featured with 128-core Nvidia GPU in Maxwell architecture, is a small developer kit for processing neural network model inference at the edge. We chose these devices as together they cover the mainstream of integrated GPUs in the market. Intel Graphics, ARM Mali GPUs, and Nvidia GPUs all have several different models and micro-architecture generations. Our solution can be deployed on other edge devices featured with similar integrated GPUs because our programming environment is unified.

We tested out the performance across a wide range of CNN models from image classification (ResNet~\cite{he2016deep}, MobileNet~\cite{howard2017mobilenets}, SqueezeNet~\cite{iandola2016squeezenet}) to object detection (SSD~\cite{liu2016ssd}, Yolo~\cite{redmon2018yolov3}). The models are retrieved from GluonCV model zoo~\footnote{\url{https://gluon-cv.mxnet.io/model_zoo/index.html}}, pre-trained using MXNet. The chosen models represent the majority of the computer vision model inference workload in the applications executing at the edge devices. These models all have multiple variants (e.g. ResNet-18, ResNet-50, etc. for ResNet) to form a model family. For the sake of space, we only evaluate our solution on one variant of each model family. Performance comparison result of one model is similar to its variants of the same family. Other poplar CV models such as VGG, Faster R-CNN and Mask R-CNN are not included in the evaluation as they are not of the interest for edge applications, occupying excessively large memory footprint and too much computational intensity.

Our solution was built on top of the code base of the TVM stack 0.5.0~\footnote{\url{https://github.com/dmlc/tvm}} with a number of improvements described in Section~\ref{sec:opt}. Via the frontend support of TVM, our solution directly consumes the GluonCV models to compile and execute. However, it is non-trivial to get the respectful baselines for these devices. For the device with Intel Graphics, we used Intel OpenVINO toolkit as the baseline, which does optimized model inference on Intel Graphics using Intel clDNN along with some graph-level optimizations. OpenVINO only restricts the support of the image classification models but not the object detection ones. For the device with ARM Mali GPU, we could not find an OpenVINO counterpart for ARM. And there is limited reliable open-sourced solution to run CNN model inference on ARM GPUs. Therefore, we had to manually conduct graph-level optimization of the models and register the operators to the corresponding ACL implementation (v19.02). While this worked, it required sophisticated programming skills, making it difficult to have the models efficiently deployed at the edge devices. Regarding Nvidia Jetson Nano, although it supports TensorRT as claimed, the official TensorRT cannot consume GluonCV models and we were not able to enable it as TensorRT is not open-sourced. Alternatively, we used MXNet (v1.4.0) backed by cuDNN (v7.0) to get the baseline numbers for Nvidia Jetson Nano. Our struggling experience in looking for the baseline results demonstrates that it is inconvenient for users to deploy CNN model inference to run on integrated GPUs at the edge using chip vendor-provided libraries.

\subsection{Overall Performance}
\label{sec:eval:overall}
As this is the model inference task, we used \emph{latency}, e.g. the processing time of \emph{one} sample, as the criterion to measure the overall performance of the list of models on various integrated GPUs. Tables~\ref{tab:openvino},~\ref{tab:acl}, and~\ref{tab:nano} detail the latency numbers of our solution (noted by ``Ours'') running on AWS DeepLens, Acer aiSage and Nvidia Jetson Nano, compared to the ones consumed by Intel OpenVINO, ACL, and cuDNN, respectively. Note that our methods do not lose any accuracy compared with baselines. The result numbers are based on the following configurations and facts:
\begin{itemize}
    \item As mentioned above that Intel OpenVINO only confined themselves to supporting image classification models, we are not able to extend the support to other models as it is not open-sourced. Therefore, both latency and speedup numbers of the object detection models in Table~\ref{tab:openvino} are omitted.
    \item Acer aiSage reduces the input size to $300\times300$ ($512\times512$ for others) due to the memory limitation of ARM Mali GPUs.
\end{itemize}

\begin{table}[tbph]
\begin{tabular}{@{}llll@{}}
\toprule
Models         & Ours (ms)     & OpenVINO (ms) & Speedup \\ \midrule
ResNet50\_v1         & 186.15  & 203.60   & 1.09    \\
MobileNet1.0      & 85.58   & 53.48    & 0.62    \\
SqueezeNet1.0     & 52.10   & 42.01    & 0.81    \\
SSD\_MobileNet1.0 & 398.48  & ---       & ---        \\
SSD\_ResNet50    & 1006.01 & ---       & ---      \\
Yolov3          & 1004.13 & ---       & ---      \\ \bottomrule
\end{tabular}
\caption{The performance comparison of our solution vs OpenVINO on AWS DeepLens. ``---'' indicates that the model is not yet supported by OpenVINO.}
\label{tab:openvino}
\vspace{-0.2in}
\end{table}

\begin{table}[tbph]
\begin{tabular}{@{}llll@{}}
\toprule
Models         & Ours (ms)     & ACL (ms)     & Speedup \\ \midrule
ResNet50\_v1         & 345.60  & 358.17   & 1.04    \\
MobileNet1.0      & 78.83   & 95.00    & 1.21    \\
SqueezeNet1.0     & 66.61  & 77.10    & 1.16    \\
SSD\_MobileNet1.0 & 243.16  & 216.87   & 0.89    \\
SSD\_ResNet50    & 777.26  & 737.90   & 0.95    \\
Yolov3          & 1097.47 & 1042.90  & 0.95    \\ \bottomrule
\end{tabular}
\caption{The performance comparison of our solution vs ACL on Acer aiSage.}
\label{tab:acl}
\vspace{-0.1in}
\end{table}

\begin{table}[tbph]
\begin{tabular}{@{}llll@{}}
\toprule
Models         & Ours (ms)    & cuDNN (ms)     & Speedup \\ \midrule
ResNet50\_v1         & 113.81  & 117.22   & 1.03    \\
MobileNet1.0      & 20.63   & 30.71    & 1.49    \\
SqueezeNet1.0     & 26.58   & 42.98    & 1.62    \\
SSD\_MobileNet1.0 & 135.5   & 197.3    & 1.47    \\
SSD\_ResNet50	   & 371.32  & 478.33   & 1.29    \\
Yolov3          & 553.79  & 802.41   & 1.45    \\
\bottomrule
\end{tabular}
\caption{The performance comparison of our solution vs cuDNN on Nvidia Jetson Nano.}
\label{tab:nano}
\vspace{-0.1in}
\end{table}

The following observations are obtainable from the numbers listed in the tables. 
\begin{itemize}
    \item Our solution in general achieves similar or better performance among the selected models across three different platforms. The slightly slower performance could potentially be improved by finer tuning algorithms and by improving the search space.
    \item Our solution for MobileNet on Intel Graphics (Deeplens) is not as good as on Mali GPUs (aiSage) or Nvidia GPUs (Nano). The reason is that our depth-wise convolution has not been fully optimized for Intel Graphics. Optimizing depth-wise convolutions on Intel Graphics using our unified IR remains our future work.
    \item For models that are popular at the edge, e.g. MobileNet and SqueezeNet, our approach outperforms both ACL and cuDNN. This is because the machine learning-based searching approach we adopted is able to figure out optimized scheduling schemes, while the chip vendor-provided libraries might not cover all convolution workloads well on all platforms.
\end{itemize}

\subsection{Optimization of Vision-specific Operators}
\label{sec:eval:vision}

\begin{table}[tbph]
\centering
\resizebox{\linewidth}{!}{
\begin{tabular}{@{}llccc@{}}
\toprule
Devices     & Models         & Before (ms)   & After (ms)    & Speedup \\ 
\midrule
\multirow{3}{*}{AWS Deeplens} &
SSD\_MobileNet1.0 & 966.20   & 398.48   & 2.42    \\ &
SSD\_ResNet50    & 1491.30  & 1006.01  & 1.48    \\ &
Yolov3          & 2610.13  & 1004.13  & 2.6      \\ 
\midrule
\multirow{3}{*}{Acer aiSage}
& SSD\_MobileNet1.0 & 1098.11   & 243.16   & 4.5     \\
& SSD\_ResNet50    & 1631.30  & 777.26   & 1.48    \\
& Yolov3          & 6429.69  & 1097.47  & 5.86     \\
\midrule
\multirow{3}{*}{Nvidia Jetson Nano}
& SSD\_MobileNet1.0 & 264   & 135.5   & 1.95     \\
& SSD\_ResNet50    & 490.4  & 371.32   & 1.32    \\
& Yolov3          & 1350  & 553.79  & 2.44     \\
\bottomrule
\end{tabular}}
\caption{The performance comparison of our solution with and without vision-specific operator optimizations on different devices.}
\label{tab:vso}
\vspace{-0.1in}
\end{table}

This subsection investigates the effectiveness of the optimization on vision-specific operators. As these operators mainly exist in object detection model, we hence only present the results for the three object detection networks here. Table~\ref{tab:vso} shows the performance difference of our solution with and without vision-specific optimization. As shown in the table, by incorporating our optimization techniques on vision-specific operators, we can achieve remarkable performance boost, e.g. up to 5.86$\times$, on all the integrated GPUs. 

The following approaches contributed to the compelling performance improvement. First, the proposed segmented sorting effectively improves loading balancing for \emph{argsort} by flattening the input array and partitioning it into smaller chunks. Second, our three-stage \emph{prefix sum} algorithm not only reduces the latency to logarithmic time, but also avoids global synchronization using register blocking. Lastly, our \emph{box\_nms} operator avoids irregular data access by aligning the most inner loop operation with threads, one step upper to blocks, the batch level to unrolled loops. And it also avoids branch divergence by initializing all output to be invalid instead of doing it in a comparison style. From Table~\ref{tab:vso}, we can observe that aiSage benefits most from the vision-specific operations. The reason is that Mali GPUs do not have shared memory in their hardware architecture, therefore, load balancing, data assessment and branch divergence matter more to the performance.

\subsection{Optimization of CONV Operators}
\label{sec:eval:conv}
This subsection inspects the performance effect improved by our optimization heuristics on convolution operators. Our implementation uses machine-learning based heuristics to perform comprehensive tuning on integrated GPUs. Table~\ref{tab:conv} shows performance improvements achieved by auto-tuning in image classification models. 

We can see from Table~\ref{tab:conv} that our auto-tuning approaches offer considerable performance improvements over the manual implementation that are not tuned. For example, our solution achieves up to $39.3\times$ and $12.78\times$ speedups for SqueezeNet on Nvidia Jetson Nano and Acer aiSage, respectively. The reason of SqueezeNet being improved the most is that the network is fairly new so there is no manually written implementation of it in good performance. There are two major contributions to the performance improvement. First, the graph tuner is able to fine-tune the data layout, hence providing the graph-level optimal schedule for each convolution kernel with various workloads and parameters. Second, AutoTVM offers more optimization opportunities at even finer granularity using machine learning based search schemes, e.g. tensor-level tuning. This imposes more architectural level optimizations, e.g. better memory access pattern, vectorization, instruction level parallelism, etc, to the convolution kernels. This is the first work that performs comprehensive machine learning-based search on integrated GPUs.

\begin{table}[tbph]
\centering
\resizebox{\linewidth}{!}{
\begin{tabular}{@{}lllll@{}}
\toprule
Devices     & Models         & Before (ms)   & After (ms)    & Speedup \\ 
\midrule
\multirow{3}{*}{AWS Deeplens}
& Resnet50\_v1     & 260     & 186.15 & 1.4    \\
& MobileNet1.0  & 558.15  & 85.58  & 6.52    \\
& SqueezeNet1.0 & 64      & 52.1   & 1.23      \\
\midrule
\multirow{3}{*}{Acer aiSage}
& Resnet50\_v1     & 727.29  & 345.6  & 2.1    \\
& MobileNet1.0  & 655.18  & 78.83  & 8.31    \\
& SqueezeNet1.0 & 1362.2  & 106.61 & 12.78      \\
\midrule
\multirow{3}{*}{Nvidia Jetson Nano}
& Resnet50\_v1     & 1088.55  & 113.81  & 9.56    \\
& MobileNet1.0  & 155.14  & 20.63  & 7.52    \\
& SqueezeNet1.0 & 1045  & 26.58 & 39.3      \\
\bottomrule
\end{tabular}}
\caption{The performance comparison of our solution with and without tuning based optimizations for convolution operators on different devices.}
\label{tab:conv}
\vspace{-0.2in}
\end{table}

\section{Related Works}
\label{sec:related}
As more and more applications deployed at the edge devices, there is active research on optimizing the deep learning model inference, especially CNN models, locally on mobile SoCs. The first place to execute the models is mobile CPUs~\cite{wu2019hpca}, which is heavily studied and optimized~\cite{liu2018optimizing}. However, as mobile CPUs typically have less FLOPs and consumes more power, it is more favorable to leverage other compute units on the SoCs to handle the deep learning model inference tasks~\cite{lane2016deepx,huynh2017deepmon}. These works focused on utilizing the mobile DSPs and GPUs to tackle the computationally-intensive operators such as matrix multiplication and convolution. However, they paid little attention to optimizing the vision-specific operators on integrated GPUs, making their model coverage normally narrow. In contrast, the solution proposed in this paper covers a wide spectrum of image classification, object detection models on mainstream integrated GPUs.

This paper uses a unified IR to lower the computation to CUDA or OpenCL codes for different devices. This idea was inherited from Halide~\cite{ragan2013halide}. We extended the implementation to handle the vision-specific operators efficiently on integrated GPUs, which by nature are not designed for this kind of control logic-intensive operations.

Heterogeneous computing for general application has been intensively studied. However, there is not much work in the deep learning literature in general, model inference in particular. Pu et al.~\cite{Pu2017TACO} extended the image processing language, Halide~\cite{ragan2013halide}, to allow users to specify which part of their applications they want to execute on hardware accelerators (FPGA in their case). Similar to their technique, our approach also allows users to offload different portions of a CNN to different devices so that programmers can quickly build and tune new CNN models. Unlike their approach, our solution offers various automatic performance tuning techniques at both graph- and tensor-level which effectively relieves programmers from tedious performance tuning for various platforms. \cite{Gambardella} proposed to use a hardware accelerator for the inference of quantized neural networks on a heterogeneous embedded system, which mainly focused on the network produced by a specific framework and the platforms with restricted resources. However, our approach is applicable to most of the models of popular frameworks. A few previous works proposed software/hardware co-design of a heterogeneous memory system for acceleration of neural networks~\cite{Azarkhish2018NeurostreamSA,Liu2018ProcessinginMemoryFE}. However, these mechanisms mainly focused on the hardware architecture design to speed up training. Our solution targets providing a unified programming framework with broader model coverage for inference.

The optimization techniques used in this paper do not change the semantics of the models so introduce no accuracy loss. There is another way to enable and expedite the deep learning model inference at the edge which either quantizes the model to reduce size~\cite{HanMD15} or customizes the model for mobile usage~\cite{howard2017mobilenets,mcdanel2017embedded,zhang2018shufflenet,tan2018mnasnet}. Essentially, it trades off some model inference accuracy to shorten the end-to-end execution time, which is beyond the scope of this paper but remains the future work.

Our solution utilized the machine learning techniques to search for satisfying schedules to execute the model inference, including the tensor-level~\cite{Chen:2018:LOT} and graph-level~\cite{liu2018optimizing}. The similar schemes were applied to other platforms as well (e.g. CPUs~\cite{liu2018optimizing}), but to the best of our knowledge, our work is the first time to search execution scheme of deep learning model inference on integrated GPUs.
\section{Conclusion}
\label{sec:concl}
This paper presents a unified solution to efficiently run CNN model inference on mainstream integrated GPUs. Our solution has a broader coverage on popular CNN models used in the computer vision domain with similar or better performance compared to the state-of-the-art solutions backed up by the high-performance libraries provided by the chip vendors. We achieve it by carefully optimizing the vision-specific operators, as well as leveraging the machine learning-based performance tuning schemes. In the meantime, our solution also allows falling back operators to execute on the accompanying CPU of the same SoC to accelerate the adoption of new models to run in high performance on edge devices. We hope our work will boost the usage of intelligent CV applications at the edge.

\section{Acknowledgment} {
The authors thank the anonymous reviewers of the paper for valuable comments. The authors are also grateful to Frank Chen and Long Gao for providing devices for experiments, and Tianqi Chen for technical assistance. The entire work was done at AWS. The first author appreciates the advice of Prof.\ John D. Owens and the funding support from the Air Force Research Lab (AFRL) and the Defense Advanced Research Projects Agency (DARPA) under agreement number FA8650-18-2-7836 when she was a grad student. 
}

\bibliographystyle{plain}
\bibliography{ref}

\end{document}